\documentclass[letterpaper]{sig-alternate-2013}

\usepackage{fdiaz}
\usepackage{subcaption}
\usepackage{amsmath}
\usepackage{algorithm}
\usepackage[noend]{algpseudocode}
\usepackage{url}

\makeatletter
\usepackage[
  pass,
]{geometry}
\def\BState{\State\hskip-\ALG@thistlm}
\makeatother

\permission{\copyright 2017 International World Wide Web Conference
  Committee \\
  (IW3C2), published under Creative Commons CC BY 4.0 License.}
\conferenceinfo{WWW 2017 Companion,}{April 3--7, 2017, Perth,
  Australia.}
\copyrightetc{ACM \the\acmcopyr}
\crdata{978-1-4503-4914-7/17/04. \\
  http://dx.doi.org/10.1145/3041021.3054197 \\
  \includegraphics{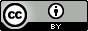}}

\newcommand{\reformulationRate}[0]{\text{RR}}
\newcommand{\satifiedClickCount}[0]{\text{SCC}}
\newcommand{\pageClickCount}[0]{\text{PCC}}
\newcommand{\gradedUtility}[0]{\text{GU}}

\begin{document}
\title{Auditing Search Engines for Differential Satisfaction Across Demographics}
\numberofauthors{6}
\author{
\alignauthor Rishabh Mehrotra\thanks{Work conducted while at Microsoft Research.}\\
\email{r.mehrotra@cs.ucl.ac.uk}\\
\affaddr{University College London}
\alignauthor Ashton Anderson\\
\email{ashton@microsoft.com}\\
\affaddr{Microsoft Research}
\alignauthor Fernando Diaz\footnotemark[1]\\
\email{diazf@acm.org}\\
\affaddr{Spotify}
\and
\alignauthor Amit Sharma\\
\email{amshar@microsoft.com}\\
\affaddr{Microsoft Research}
\alignauthor Hanna Wallach\\
\email{wallach@microsoft.com}\\
\affaddr{Microsoft Research}
\alignauthor Emine Yilmaz\\
\email{emine.yilmaz@ucl.ac.uk}\\
\affaddr{University College London}
}

\maketitle
\begin{abstract}
  Many online services, such as search engines, social media
  platforms, and digital marketplaces, are advertised as being
  available to any user, regardless of their age, gender, or other
  demographic factors. However, there are growing concerns that these
  services may systematically underserve some groups of users. In this
  paper, we present a framework for internally auditing such services
  for differences in user satisfaction across demographic groups,
  using search engines as a case study. We first explain the pitfalls
  of na\"{\i}vely comparing the behavioral metrics that are commonly
  used to evaluate search engines. We then propose three methods for
  measuring latent differences in user satisfaction from observed
  differences in evaluation metrics. To develop these methods, we drew
  on ideas from the causal inference literature and the multilevel
  modeling literature. Our framework is broadly applicable to other
  online services, and provides general insight into interpreting
  their evaluation metrics.
\end{abstract}

\vspace{-1mm}

\section{Introduction}
\label{sec:introduction}

Modern search engines are complex, relying heavily on machine learning
methods to optimize search results for user satisfaction. Although
machine learning can address many challenges in web search, there is
also increasing evidence that suggests that these methods may
systematically and inconspicuously underserve some groups of
users~\cite{bolukbasi:word2vec-bias,angwin:machine-bias}. From a
social perspective, this is troubling. Search engines are a modern
analog of libraries and should therefore provide equal access to
information, irrespective of users' demographic
factors~\cite{ifla:code-of-ethics}. Even beyond ethical arguments,
there are practical reasons to provide equal access. From a business
perspective, equal access helps search engines attract a large and
diverse population of users. From a public-relations perspective,
service providers and the decisions made by their services are under
increasing scrutiny by
journalists~\cite{diakopoulos:algorithmic-accountability} and
civil-rights
enforcement~\cite{whitehouse:big-data-civil-rights,barrocas:big-data-disparate}
for seemingly unfair behavior.

One way to assess whether a search engine provides equal access is to
look for differences in user satisfaction across demographic
groups. If users from one group are consistently less satisfied than
users from another, then these users are likely not being provided
with equal search experiences. However, in practice, measuring
differences in satisfaction is non-trivial. One demographic group may
issue very different queries than another. Or, two groups may issue
similar queries, but with different intents. Any differences in
aggregate evaluation metrics will therefore reflect these contextual
differences, as well as any differences in user
satisfaction. Moreover, since user satisfaction cannot be measured at
scale using explicit feedback, search engines often rely on implicit
feedback based on behavioral signals, such as the number of clicks or
the dwell time (i.e., the time spent on a
page)~\cite{white2016interaction}. Even controlling for differences in
the types of queries issued and in user intents, these signals may
themselves be systematically influenced by demographics. Therefore, we
cannot interpret evaluation metrics based on them as being direct
reflections of user satisfaction. For example, if older users
typically read more slowly than younger users, then a metric based on
dwell time will, on average, be higher for older users, regardless of
their levels of satisfaction.

To better understand the subtleties of this challenge, consider a
search engine with users that span a wide range of age groups. As an
example, suppose that users in their twenties comprise 80\% of the
search traffic, while users over the age of fifty comprise 10\% of the
search traffic. Suppose also that older users pose many more queries
about retirement planning compared to younger users. Finally, suppose
that the search engine relies on the dwell time for clicked results to
measure user satisfaction~\cite{kim2014modeling}. It might seem
natural to consider the average value of this metric in order to make
product decisions. However, simply considering the average value of
the metric across all users will underemphasize the effectiveness of
the search engine on retirement planning queries. Moreover, if the
search engine ranks documents poorly for retirement planning queries,
then older users' low levels of satisfaction will be obscured. Even
considering retirement planning queries in isolation, the average
value may overemphasize the satisfaction of younger users focused on
early-career retirement planning, again obscuring low levels of
satisfaction for older users focused on late-career retirement
planning. Finally, if the metric was calibrated with respect to
younger users, then a dwell time of 30 seconds may be sufficient to
demonstrate satisfaction. But if older users read more slowly, then a
30-second threshold might result in overoptimistic satisfaction
estimates for these users.

In this paper, we propose three methods for measuring latent
differences in user satisfaction from observed differences in
evaluation metrics. All three methods are internal auditing
methods---i.e., they use internal system information.

Our first two methods aim to disentangle user satisfaction from other
demographic-specific variation by controlling for the effects of
demographic factors on behavioral metrics; if we can recover an
estimate of user satisfaction for each metric and demographic group
pairing, then we can compare these estimates across groups. Any
auditing method must strike a balance between generalizability and
controlling for as many confounding factors as possible: the more
controls in place, the less generalizable the conclusions. Our first
method, context matching, controls for two confounding contextual
differences: the query itself and the intent of the user
(section~\ref{sec:context}). Because this method attempts to match
users' search contexts as closely as possible, it can only be applied
to a restricted set of queries. Our second method is a multilevel
model for the effect of query difficulty on evaluation metrics
(section~\ref{sec:multilevel}). In contrast, this method controls for
fewer confounding factors, but is more generalizable.

Although these methods shed light on observed differences between
demographic groups, they say little about the substantive question of
differential satisfaction across demographic groups. For our third
method, we therefore take a different approach. Instead of estimating
user satisfaction for each demographic group and then comparing these
estimates, we estimate the latent differences directly
(section~\ref{sec:pairwise-sat}). Because we are not interested in
absolute levels of satisfaction, this is a more direct way to achieve
our goal. This method infers which impression, among a randomly
selected pair of impressions, led to greater user satisfaction. Then,
using our second method, we set a threshold for differences that are
so large that they are unlikely to explained by anything other than
genuine differences in user satisfaction.

We used all three methods to audit Bing---a major search
engine---using proprietary data focusing specifically on age and
gender. We found significant differences in raw usage patterns and
aggregate evaluation metrics for different demographic groups
(section~\ref{sec:demographic}). However, after using our methods to
control for confounding contextual differences, we found much less
variation across groups
(sections~\ref{sec:context},~\ref{sec:multilevel},
and~\ref{sec:pairwise-sat}). Overall, we found no difference in
satisfaction between male and female users, but we did find that older
users appear to be slightly more satisfied than younger users.

Finally, for comparison, we also used our third method to conduct an
external audit of a leading competitor to Bing using publicly
available data from comScore (section~\ref{sec:external}).

\section{Related Work}

In this section, we briefly survey related work in three distinct
research areas: fairness in machine learning, demographics and web
search, and user satisfaction in web search.

\subsection{Fairness in Machine Learning}
As we described in the previous section, all three of our methods are
internal auditing methods. Internal auditing methods are employed by
service providers to review their own services using internal system
information. For example, a social media platform might audit its own
algorithms by examining the actual decisions made for the true
population of users, with demographic attributes revealed. As another
example, Feldman et al. presented a method for detecting and
correcting potential demographic biases in training
data~\cite{feldman:disparate-impact}. External auditing methods differ
from internal auditing methods in that they rely only on publicly
available information. External auditing methods are typically
employed by third parties to review services without the cooperation
of the service providers. As a result, external auditing methods
cannot rely on internal system information. For example, Adler et
al. presented a method for detecting biased decisions by probing an
API~\cite{adler:black-box-auditing}. Other work sought to detect
unfairness through reverse A/B testing with synthetic
users~\cite{lecuyer:sunlight,datta:algorithmic-transparency},
algorithmic auditing~\cite{sandvig2014auditing}, and analysis of
predictions made by supervised machine learning
methods~\cite{hardt2016equality}.\looseness=-1

To the best of our knowledge, all previous work on internal and
external auditing methods assumes that indicators of effectiveness are
directly observable (e.g., accuracy). In contrast, our focus is on
scenarios where the evaluation metrics themselves may be influenced by
confounding demographic factors. Disentangling effectiveness from
other demographic-specific variation is crucial in these scenarios.

\subsection{Demographics and Web Search}

Researchers have studied the behavior of search engine users in
various settings: Ford et al.~\cite{ford:demographics-search}
conducted a controlled experiment involving masters students, varying
age and gender; Weber and Castillo~\cite{weber:search-demographics,
  weber2011uses} studied differences in user behavior using search log
data; Bi et al.~\cite{bi:inferring-social-from-web} demonstrated that
search behavior can be used to predict demographic attributes; Lorigo
et al.~\cite{lorigo:task-gender-search-evaluation} studied the effect
of gender on user behavior and found a relationship between gender and
eye gaze patterns; and several
studies~\cite{large:gender-search-school,roy:gender-search-school}
have established that the search behavior of school-aged children
varies by gender. Other related studies measured the impact of
demographics on search results~\cite{hannak2013measuring}, examined
the search engine manipulation effect~\cite{epstein2015search},
explored demographic context as a means to improve search results for
ambiguous queries, and analyzed gender differences in search
perceptions~\cite{zhou2014gender}. Together, these studies ground the
role of demographics in evaluating search engines and motivate our
work.\looseness=-1

\subsection{User Satisfaction in Web Search}

Although search engines are often evaluated using metrics based on
behavioral signals, several studies have suggested that these metrics
are sensitive to a variety of factors: Hassan and
White~\cite{hassan:personalized-metrics} demonstrated that evaluation
metric values vary dramatically by user; Carterette et
al.~\cite{carterette:cikm2012} made a similar observation and
therefore incorporated user variability into evaluation metrics; and
Borisov et al. studied the degree to which metrics are sensitive to a
user's search context~\cite{borisov-context-aware-2016}. Our work
adopts a similar philosophy, focusing on measuring the extent to which
demographics affect metrics.

\section{Data and Metrics}
\label{sec:data}

We selected a random subset of Bing's desktop and laptop users from
the English-speaking US market, and focused on their log data from a
two week period during February, 2016. We removed spam using standard
bot-filtering methods, and discarded queries that were not manually
entered. By performing these filtering steps, we could be sure that
any differences in evaluation metrics were not due to differences in
devices, languages, countries, or query input methods.

We enriched these data with user demographics, focusing on
self-reported age and (binary) gender information obtained during
account registration. We discarded data from any users older than 74,
and binned the remaining users according to generational boundaries:
(1) younger than 18 (post-millennial), (2) 18--34 (millennial), (3)
35--54 (generation X), and (4) 55--74 (baby
boomers).\footnote{\url{http://www.pewresearch.org/fact-tank/2016/04/25/millennials-overtake-baby-boomers/}}
To validate each user's self report, we predicted their age and gender
from their search history, following the approach of Bi et
al.~\cite{bi:inferring-social-from-web}. We then compared their
predicted age and gender to their self-reported age and gender. If our
prediction did not match their self report, we discarded their
data. Approximately 51\% of the remaining users were male. In
contrast, the distribution of users across the four age groups was
much less even, with the younger age groups containing substantially
fewer users ($<$1\% and 13\% for post-millennial and millennial,
respectively) compared to the older age groups (41\% and 45\% for
generation X and baby boomers, respectively).

Finally, we labeled the remaining queries with topic information,
using the approach of Bennett et
al.~\cite{bennett:web-page-classifier}.  For each query, we
categorized the top three results, as well as all of the results
clicked on by users, into the top two levels of the Open Directory
Project\footnote{\url{https://www.dmoz.org/}} topic hierarchy using a
state-of-the-art text-based classifier. We then selected the most
common topic from the categories predicted for that query.

After these steps, we were left with 32 million search impressions,
involving 16 million distinct queries. (A search impression is a
unique view of a results page presented to a user in response to a
query.) These queries were issued by 4 million distinct users over 17
million distinct sessions.

Search engines log massive amounts of user interaction data that are
retrospectively analyzed to detect, design, and validate behavioral
signals that could serve as an implicit source of feedback. As
described in the previous section, evaluation metrics based on these
behavioral signals are often used as a proxy for user
satisfaction. Drawing upon previous work, we considered four different
evaluation metrics, each intended to operationalize user satisfaction:
graded utility, reformulation rate, page click count, and successful
click count. For graded utility, page click count, and successful
click count, higher values mean higher user satisfaction; for
reformulation rate, the relationship is reversed.

Graded utility is a model-based metric that provides a four-level
estimate of user satisfaction based on search outcome and user effort.
Jiang and Hassan~\cite{jiang:query-utility} demonstrated
experimentally that graded utility can predict subtle changes in
satisfaction more accurately than other state-of-the-art methods,
affording greater insight into search satisfaction.

Reformulation rate is the fraction of queries that were followed by
another, reformulated, query. Reformulating a query is generally a
strong indication that a user is dissatisfied with the search results
for their original query. Hassan~\cite{hassan2013beyond} showed that
evaluation metrics based on reformulation rate successfully predict
query-level satisfaction.

Page click count is the the total number of clicks made by a user on a
results page. This evaluation metric is thought to reflect the user's
level of engagement with the results page.

Although click-based evaluation metrics, such as page click count,
have traditionally been used to measure user satisfaction, more
engagement does not always correspond to higher
satisfaction. Researchers have therefore proposed time-based metrics
that are often more robustly correlated with user
satisfaction. Successful click count is the number of clicked results
with dwell times longer than 30 seconds~\cite{buscher2009segment}.

\section{Demographic Differences}
\label{sec:demographic}

In this section, we describe observed differences between demographic
groups. We focus on differences in the types of queries issued by
users and differences in evaluation metrics.

\begin{figure}
	\begin{center}
       \includegraphics[width=3.25in]{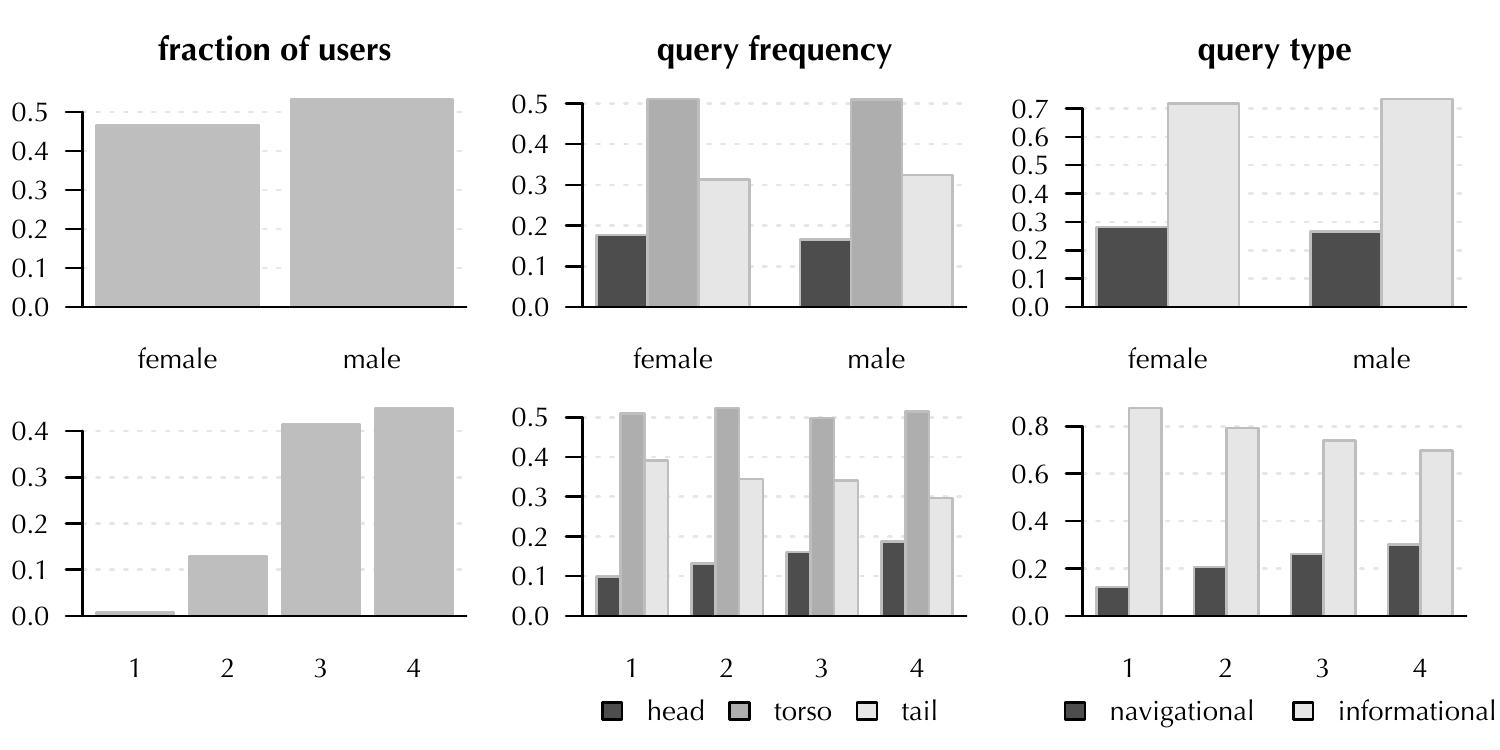}
	\end{center}
	\caption{Raw usage patterns for queries issued by users with
          different genders (top) and age groups
          (bottom).}\label{fig:characteristics}
\end{figure}

\subsection{Differences in Queries}
\label{sec:demographic:query}

First, we found that users from different demographic groups issued
different types of queries (figure \ref{fig:characteristics}). As
described in the previous section, roughly half of the users were
male. However, a higher proportion of female users (28\%) issued
navigational queries compared to male users (26\%). Although similar
proportions of male and female users ($\sim$17\%) issued head queries,
slightly more male users issued tail queries. (The top 20\% and bottom
30\% of queries by search traffic are called head and tail queries,
respectively.) Based on these differences alone, we would expect male
users to exhibit worse values for the evaluation metrics described in
the previous section. In contrast to gender, the distribution of users
across the four age groups was much less even, with the younger age
groups containing substantially fewer users than the older age
groups. We found that a higher proportion of older users (30\%) issued
navigational queries compared to younger users (13\%), while younger
users (39\%) were more likely to issue tail queries compared to older
users (30\%).\looseness=-1

We also compared the actual queries issued by users from different
demographic groups. Specifically, we computed the Kullback--Leibler
divergence between pairs of (smoothed) distributions over queries
issued by different age groups. The queries issued by the youngest age
group were most similar to the second-youngest age group
($\textrm{D}_{12} = 0.0385$) and least similar to the two oldest age
groups ($\text{D}_{13}=0.0415$, $\text{D}_{14}=0.0480$). We observed
the same pattern for the other age groups, suggesting that users who
are close in age are more likely to issue similar queries than users
whose ages are further apart. Given the uneven distribution of users
across age groups, we therefore hypothesize that evaluation metrics
will be skewed to reflect topics queried by older users, potentially
overlooking topics queried by younger users.


\begin{figure}
  \centering
  \begin{subfigure}{.23\textwidth}
    \centering
    \includegraphics[width=1.5in]{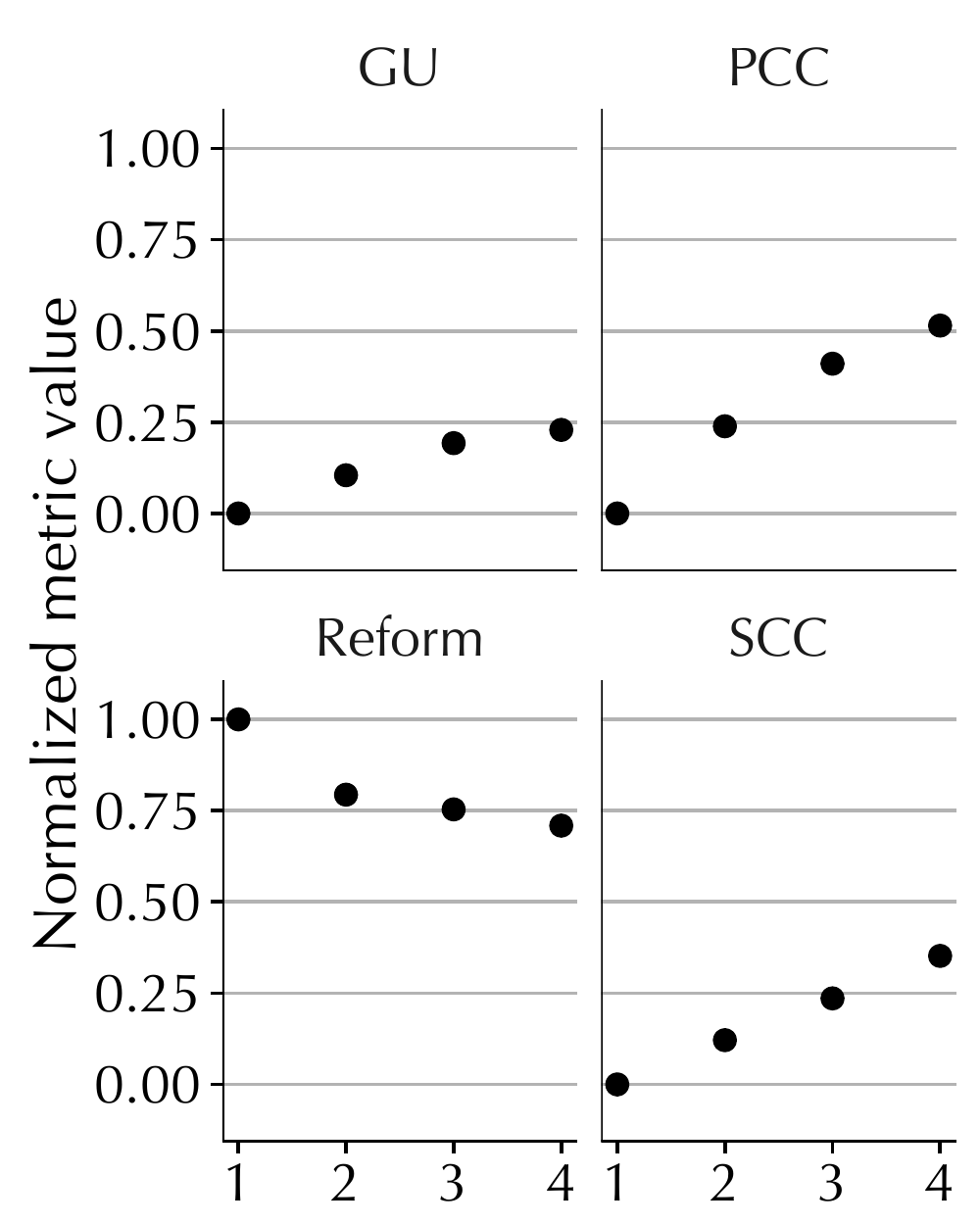}
    \caption{age\label{fig:overall_age}}
  \end{subfigure}
  \hfill
  \begin{subfigure}{.23\textwidth}
    \centering
    \includegraphics[width=1.5in]{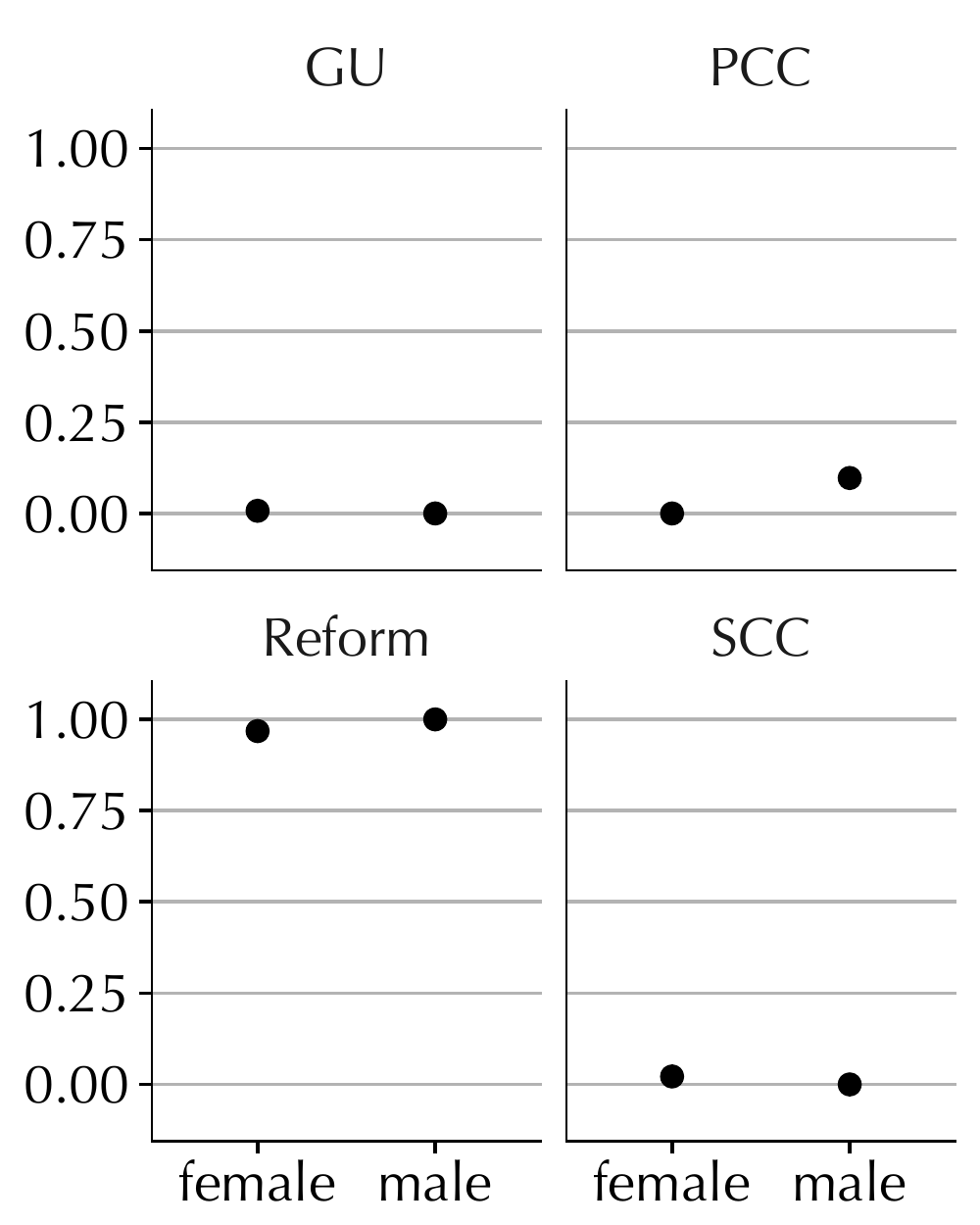}
    \caption{gender\label{fig:overall_gender}}
  \end{subfigure}
  \caption{Raw normalized query-averaged values for each metric by age
    groups (a) and genders (b). ``GU'' denotes graded utility; ``PCC''
    denotes page click count; ``Reform'' denotes reformulation rate;
    ``SCC'' denotes successful click count. Error bars (one standard
    error) are present in all plots, but are mostly so small that they
    cannot be seen.}
  \label{fig:overall}
  \vspace{-1mm}
\end{figure}

\subsection{Differences in Evaluation Metrics}
\label{sec:demographic:performance}

Next, we compared the evaluation metrics described in
section~\ref{sec:data} across demographic groups, without controlling
for any confounding demographic-specific variation. For each metric
and demographic group pairing (e.g., graded utility and millennial),
we computed the average metric value for each query issued by that
group (by averaging over impressions) and then averaged these
values. By computing query-averaged values, we ensured that our
results were not dominated by the most popular queries. Finally, we
normalized the query-averaged values to lie between zero and one. To
do this, we identified the minimum and maximum values for each metric
over the demographic groups, subtracted the corresponding minimum off
of each value, and divided each result by the corresponding maximum
minus the minimum.

We provide the normalized query-averaged values for each metric and
age group pairing in figure~\ref{fig:overall_age}. The metrics all
follow the same trend: older users have better values (lower for
reformulation rate, higher for the other metrics) than younger
users. Furthermore, the differences are quite large---for example,
users in the youngest age group have a normalized query-averaged
successful click count value of zero, while users in the oldest age
group have a value of 0.31. However, as described in
section~\ref{sec:introduction}, we cannot conclude that this trend
means that older users are genuinely more satisfied than younger
users; these differences may be due to other demographic-specific
variation. For example, users from different age groups issued
different types of queries, so this trend may simply reflect this
contextual difference.

In figure~\ref{fig:overall_gender}, we provide the normalized
query-averaged values for each metric and gender pairing. In contrast
to age, there do not appear to be any differences between
genders. Although this finding is reassuring, we cannot conclude that
it means that male and female users are equally satisfied; there may
be a large difference in satisfaction that is canceled out by other
demographic-specific variation.
\section{Context Matching}
\label{sec:context}

There are many possible sources of demographic-specific variation that
could explain the results in the previous section, some of which may
be difficult to observe and thus to control for. However, one obvious
possibility is that the observed differences in evaluation metrics
between demographic groups are due to differences in the types of
queries issued. For example, if younger users issue harder queries
than older users, then this could explain their lower values for the
evaluation metrics. In this section, we present our first method for
disentangling user satisfaction from other demographic-specific
variation. This method recovers an estimate of user satisfaction for
each metric and demographic group pairing by controlling for two
confounding contextual differences: the query itself and the intent of
the user.

We drew on well-established ideas from the causal inference literature
to develop a matching method similar to those used in medicine and the
social sciences~\cite{rubin2006matched}. Specifically, for each
demographic factor (i.e., age or gender), we made sure that the
impressions from that factor's groups were as close to identical as
possible. By focusing on near-identical contexts, we were able to
control for as many sources of demographic-specific variation as
possible. To do this, we used several filtering steps, each of which
was selected to minimize the chance that any observed differences in
evaluation metrics between demographic groups were due to anything
other than genuine differences in user satisfaction.

We first restricted the data to navigational queries because they are
generally less ambiguous than informational
queries~\cite{wang2010query}. We then retained only those queries with
at least ten impressions from each demographic group. To control for
the intent of the user, we followed the approach of Radlinski et
al.~\cite{radlinski2010inferring}. Specifically, for each query, we
identified the search result with the most final successful clicks. (A
final successful click is a successful click---i.e., a click with a
dwell time longer than 30 seconds---that terminates the query.) We
then discarded any impression whose final successful click was not on
that result. Finally, to be certain that the users had the same
choices available to them when making those clicks, we kept only those
impressions with the same results page (up to the first eight
results). After these steps, we were left with 1.2 million
impressions, involving 19,000 distinct queries, issued by 617,000
distinct users.

Following the approach described in section~\ref{sec:demographic}, for
each metric and demographic group pairing, we computed the average
metric value for each query issued by that group (by averaging over
impressions) and then averaged these values. By considering only the
1.2 million impressions described in the previous paragraph, we could
be sure we were comparing impressions that were for the same query
with the same results page and which resulted in the same search
result being the final successful click---a proxy for the intent of
the user. That said, our filtering steps did not allow us to control
for more subtle sources of demographic-specific variation.

\begin{figure}
    \centering
    \begin{subfigure}{.23\textwidth}
        \centering
        \includegraphics[width=1.5in]{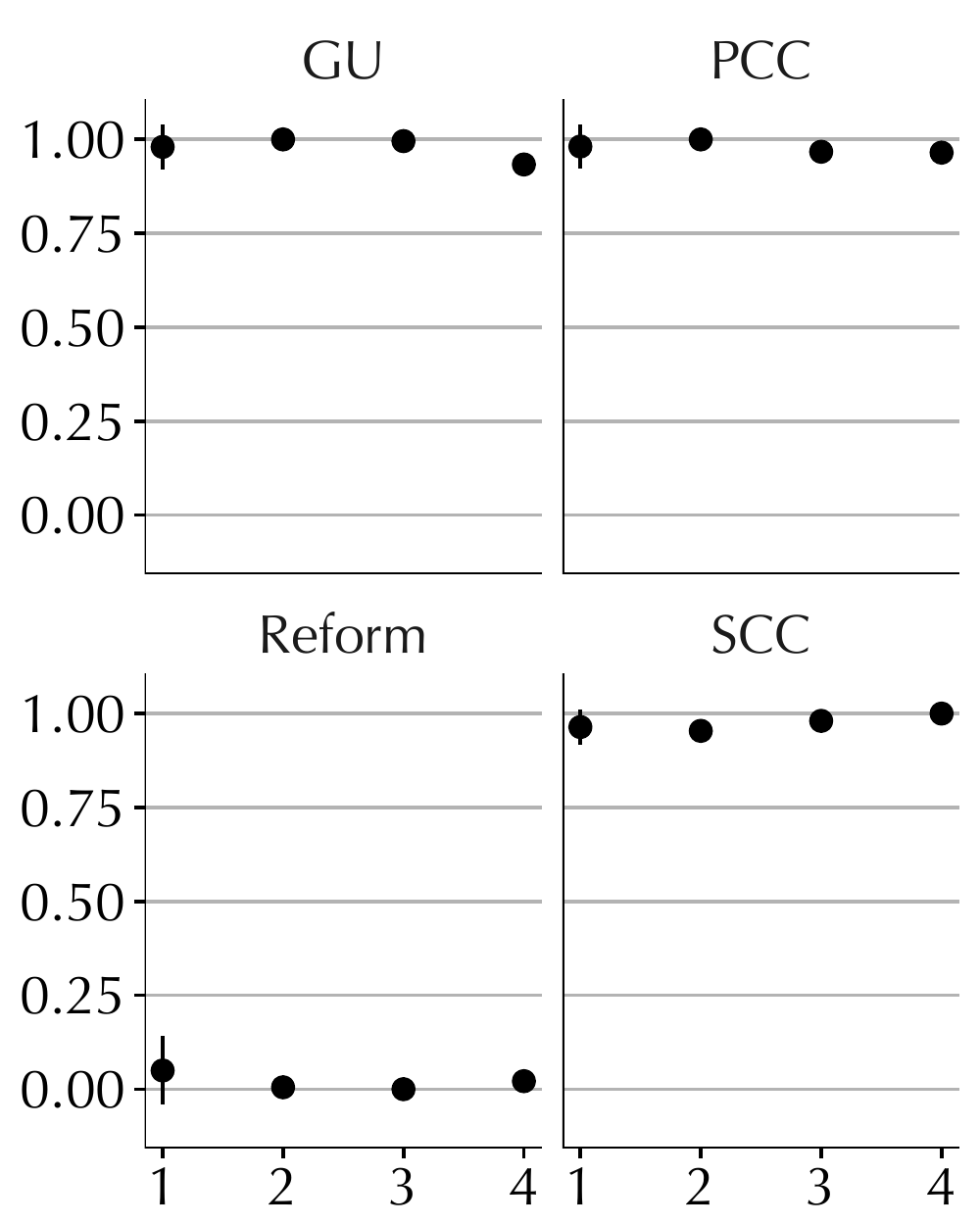}
        \caption{age}
        \label{fig:serp_age}
    \end{subfigure}%
    \hfill
    \begin{subfigure}{.23\textwidth}
        \centering
        \includegraphics[width=1.5in]{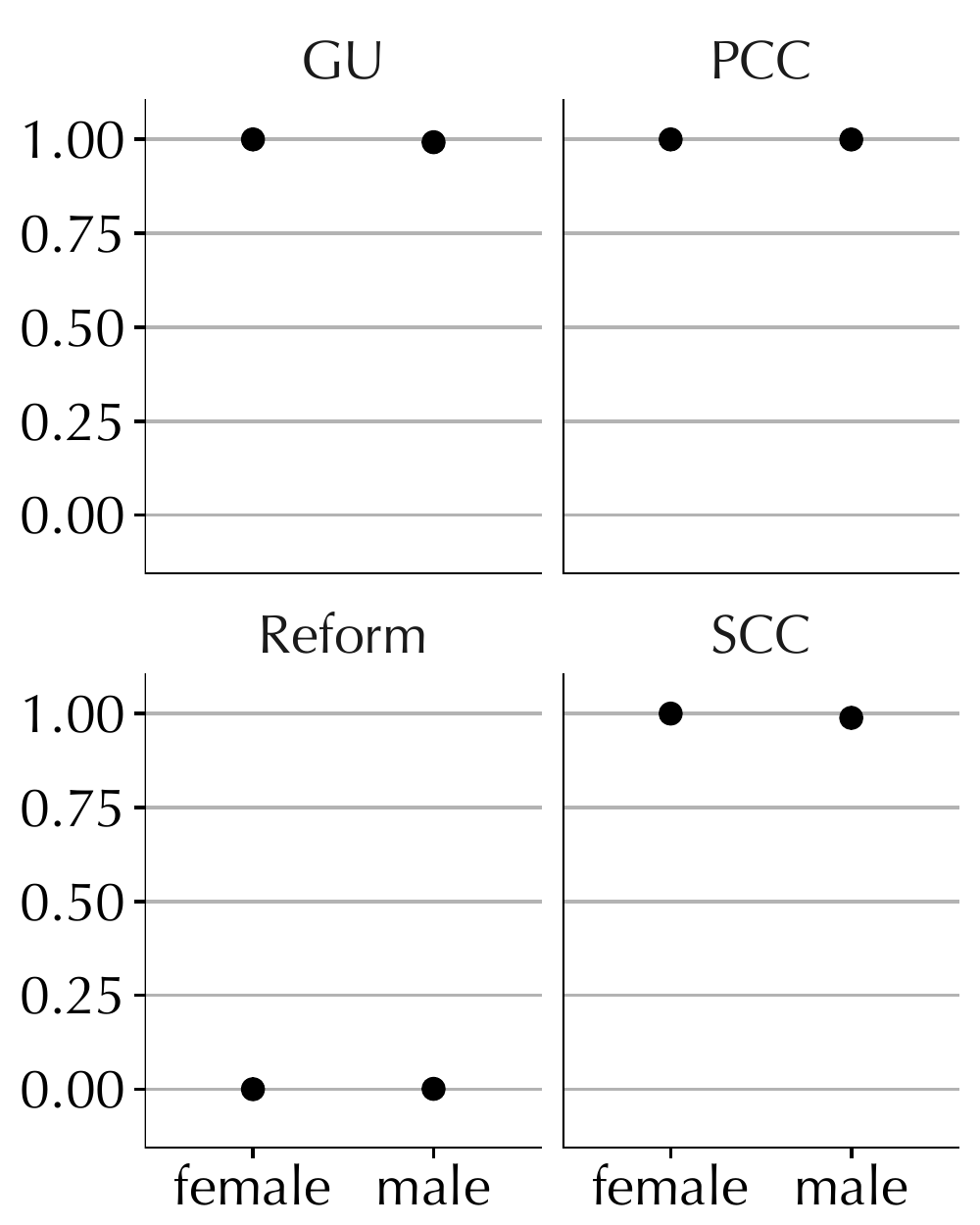}
        \caption{gender}
        \label{fig:serp_gender}
    \end{subfigure}
  \caption{Context-matched normalized query-averaged values for each
    metric by age groups (a) and genders (b). ``GU'' denotes graded
    utility; ``PCC'' denotes page click count; ``Reform'' denotes
    reformulation rate; ``SCC'' denotes successful click count. Error
    bars (one standard error) are present in all plots, but are mostly
    so small that they cannot be seen.}
    \label{fig:serp}
\end{figure}

In figure~\ref{fig:serp_age}, we provide normalized query-averaged
values for each metric and age group pairing, computed using the
context-matched data. There is much less variation across age groups
than in figure~\ref{fig:overall_age} (all data). This finding suggests
that the trend described in the previous section is unlikely to be due
to genuine differences in user satisfaction. If the search engine were
systematically underserving younger users, we would expect to see the
same trend in figure~\ref{fig:serp_age}. In
figure~\ref{fig:serp_gender}, we provide the normalized query-averaged
values for each metric and gender pairing. As in
figure~\ref{fig:overall_gender}, there do not appear to be any
differences between genders. The similarity of these figures means
that we can be reasonably confident that male and female users are
equally satisfied.\footnote{We still cannot be completely sure because
  there still may be a large difference in satisfaction that is
  canceled out by other demographic-specific variation; however,
  because we considered impressions that were for the same query with
  the same results page and which resulted in the same search result
  being the final successful click, this is very unlikely.}

We note that the normalized query-averaged values computed using the
context-matched data are significantly better (lower for reformulation
rate, higher for the other metrics) than the values in
section~\ref{sec:demographic}. This is likely because our filtering
steps restricted the data to queries that were popular enough to be
issued ten or more times by each demographic group and that led to
consistent result pages across impressions. Such queries are mostly
head queries, which are generally associated with higher levels of
satisfaction~\cite{downey2007heads}.

As we described in section~\ref{sec:introduction}, any auditing method
must strike a balance between generalizability and controlling for as
many confounding factors as possible. Although our context-matching
method requires a restricted data set, making it less generalizable,
it controls for both the query and the intent of the user, leading to
a more reliable estimate of user satisfaction for each metric and
demographic group pairing. Moreover, in many scenarios, focusing on
only the most popular queries is a very reasonable thing to do,
especially if these queries account for the majority of impressions.

\section{Multilevel Modeling}
\label{sec:multilevel}

In this section, we present our second method for disentangling user
satisfaction from other demographic-specific variation. Like our
context-matching method, this method recovers an estimate of user
satisfaction by controlling for confounding contextual differences;
however, it only controls for characteristics of the query itself and
not for the intent of the user. As a result, this method is more
generalizable, and we were able to use it without restricting the
data.

We drew on the multilevel modeling literature~\cite{gelman2006data} to
develop a new statistical model for the effect of query difficulty on
evaluation metrics, controlling for the topic of the query and
demographics of the user who issued the query. We then used this model
to examine the effects of age and gender on each of the four
evaluation metrics described in section~\ref{sec:data}, for fixed
query difficulties and topics. Because the model does not control for
the intent of the user, we cannot be sure that these effects are due
to differences in user satisfaction.

The model operationalizes the following intuition: we expect that
queries with different difficulties will lead to different metric
values. We also expect that queries about different topics will lead
to different metric values, as will queries issued by users with
different demographics. The model uses two levels to capture this
intuition: the first level accounts for differences across age,
gender, and topic combinations, while the second level models the
differences themselves.

Letting $Y_i$ denote the value of one of the four evaluation metrics
described in section~\ref{sec:data} (i.e., graded utility,
reformulation rate, page click count, or successful click count) for
the $i^{\textrm{th}}$ impression in our data set, the model assumes
that
\begin{equation}
  \label{eqn:multilevel}
  \mathbb{E}[Y_i] = f^{-1}(\alpha_{a_i g_i t_i} + \beta_{a_i g_i t_i}
  X_i),
\end{equation}
where $f(\cdot)$ is a link function; $a_i$ and $g_i$ are the age and
gender of the $i^{\textrm{th}}$ impression's user; and $t_i$ and $X_i$
are the topic and difficulty of the $i^{\textrm{th}}$ impression's
query. We can interpret $\alpha_{a_i g_i t_i}$ and $\beta_{a_i g_i
  t_i}$ as the intercept and slope, respectively, of $Y_i$ with
respect to $X_i$. The model has a different intercept and slope for
each unique age, gender, and topic combination $a \times g \times
t$. Therefore, if $a_i = a$, where $a \in \{1,2,3,4\}$, $g_i = g$,
where $g \in \{ \textrm{male}, \textrm{female}\}$, and $t_i = t$,
where $t$ is a unique topic, then $\alpha_{a_i g_i t_i} = \alpha_{a g
  t}$ and, similarly, $\beta_{a_i g_i t_i} = \beta_{a g t}$.

At the second level, the model further assumes that $\alpha_{a g t}$
and $\beta_{a g t}$ are a linear combinations of age, gender, and
topic indicator variables, as well as corresponding interaction
terms:
\begin{equation}
  \begin{pmatrix} \alpha_{a g t}\\ \beta_{a g t}
  \end{pmatrix} = \begin{pmatrix}\mu_0\\\mu_1\end{pmatrix}
  + \begin{pmatrix}\alpha_a\\\beta_a\end{pmatrix}
    + \begin{pmatrix}\alpha_g\\\beta_g\end{pmatrix}
      + \begin{pmatrix}\alpha_t\\\beta_t\end{pmatrix}
        + \begin{pmatrix}\alpha_{a \times g \times t}\\\beta_{a \times
            g \times t}\end{pmatrix}.
\end{equation}
Finally, the model assumes that the coefficients at the second level
are drawn from a mean-zero Gaussian distribution:
\begin{equation}
  \begin{pmatrix}\alpha_k\\\beta_k\end{pmatrix} \sim
    \mathcal{N}\left( \begin{pmatrix}0\\0\end{pmatrix},
      \Sigma_k\right) \textrm{ where } k \in \{a,g,t\}.
\end{equation}

To estimate the difficulty of each query, we sorted the queries issued
by each demographic group according to their graded utility values. We
then averaged each query's percentile positions in these lists to
obtain an estimate of its difficulty that is uncorrelated with the
demographics of the users who issued it. Most methods for estimating
the difficulty of a query are based on behavioral signals, such as the
reformulation rate or the dwell
time~\cite{feild:frustration,white:switching}. Because behavioral
signals may themselves be systematically influenced by demographics,
we were unable to use these methods.

%

We used a random sample of 1.4 million impressions to fit a different
version of the model for each evaluation metric. Because graded
utility ranges from negative one to positive one, we used a Gaussian
model with an identity link function; because reformulation rate
ranges from zero to one, we used a binomial model with a logit link
function; and, because page click count and successful click count are
both non-negative integers, we used a Poisson model with a logarithmic
link function. We fit each evaluation metric's version of the model
using Bayesian inference techniques~\cite{gelman2014bayesian}.

\begin{figure*}
\centering
 \begin{subfigure}{.48\textwidth}
   \centering
\includegraphics[width=3.4in]{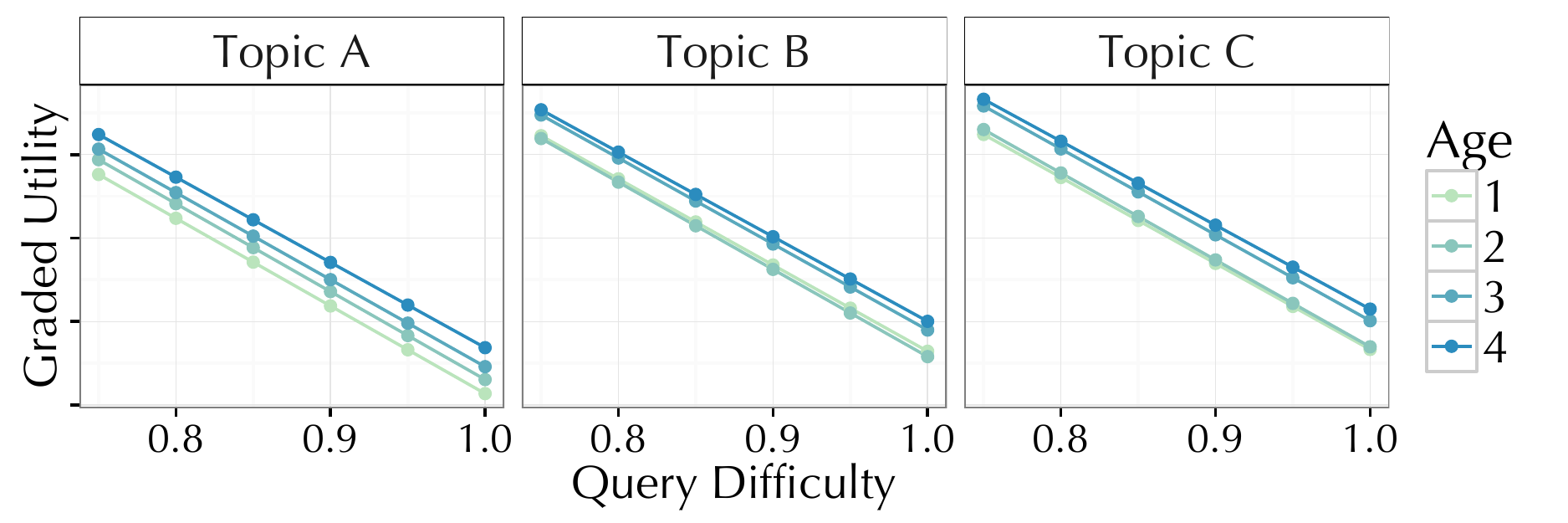}
 \caption{graded utility (hardest queries)}
\label{fig:metpred-qu}
 \end{subfigure}%
 \begin{subfigure}{.48\textwidth}
   \centering
\includegraphics[width=3.4in]{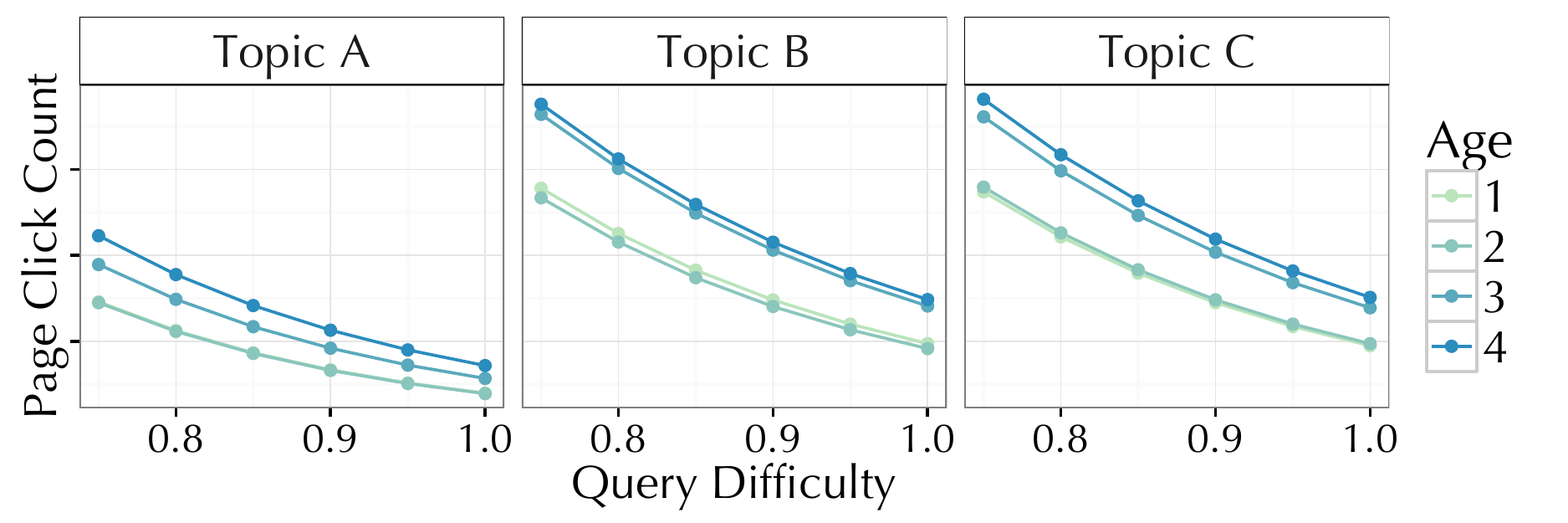}
\caption{page click count (hardest queries)}
\label{fig:metpred-pcc-hard}
\end{subfigure}
 \begin{subfigure}{.48\textwidth}
   \centering
\includegraphics[width=3.4in]{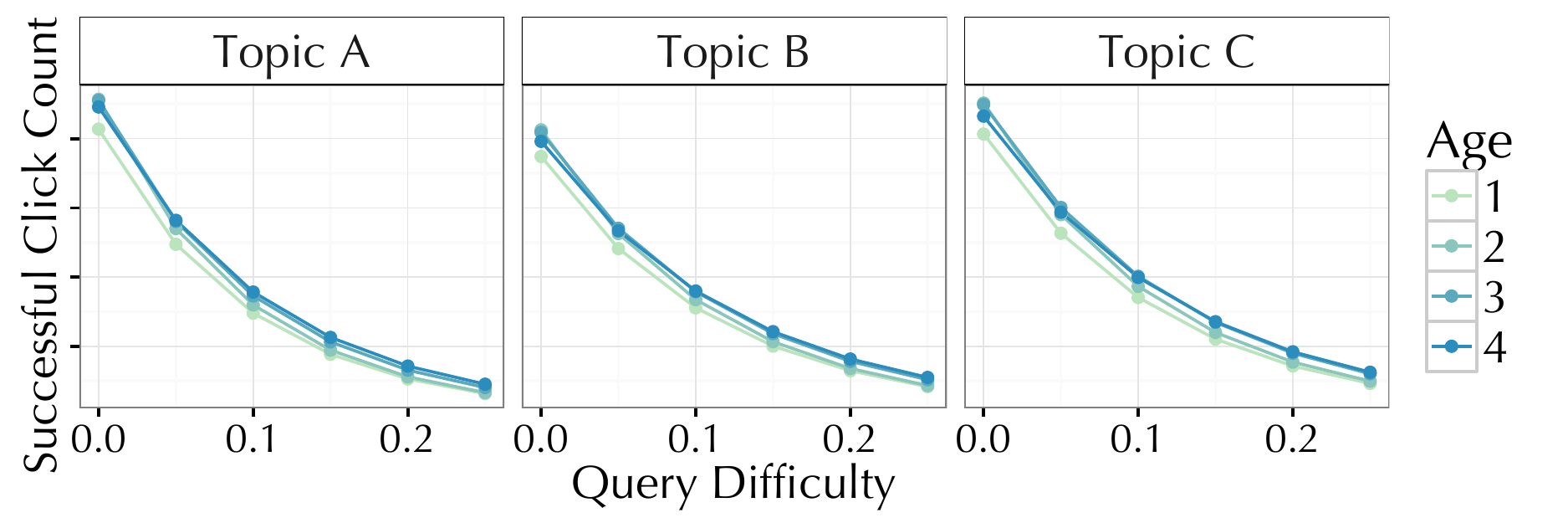}
 \caption{successful click count (easiest queries)}
\label{fig:metpred-scc-easy}
\end{subfigure}
\begin{subfigure}{.48\textwidth}
  \centering
\includegraphics[width=3.4in]{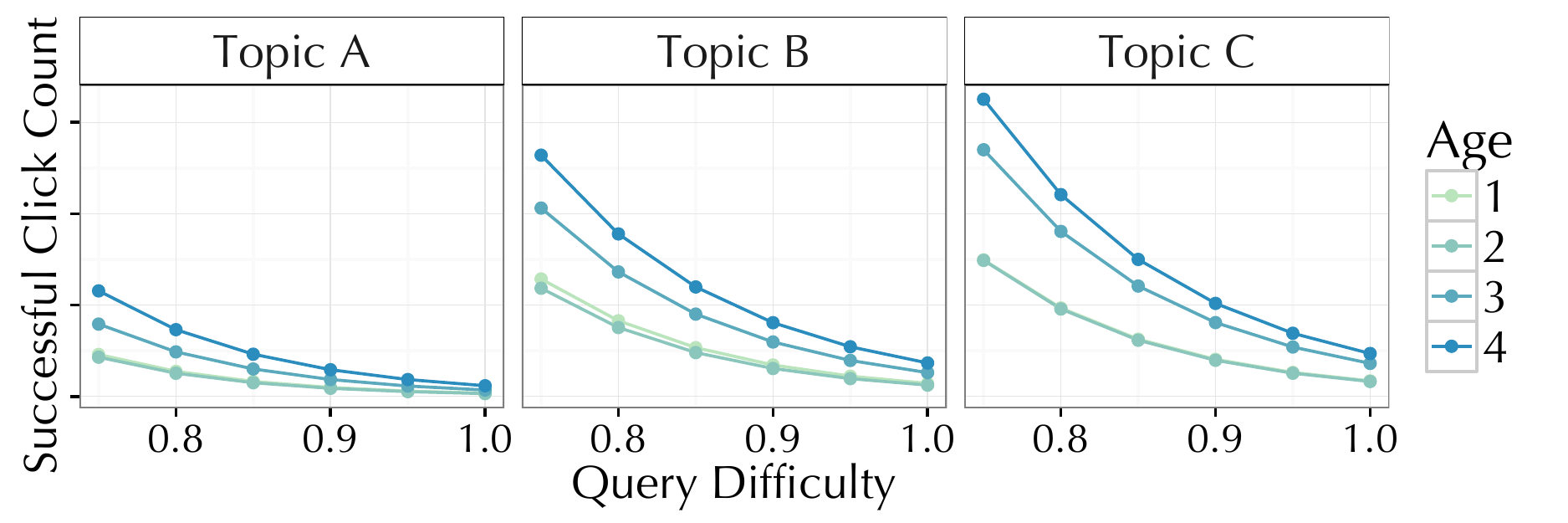}
 \caption{successful click count (hardest queries)}
\label{fig:metpred-scc-hard}
 \end{subfigure}%
\caption{Evaluation metrics according to our model. For graded utility
  (a) and page click count (b), we show only the six hardest query
  difficulties; for successful click count, we show the six easiest
  (c) and six hardest (d) query difficulties.}
\label{fig:permetrics}
\end{figure*}

Overall, we found varying levels of satisfaction across different
topics. We also found that satisfaction decreased with query
difficulty. Again, we found that gender had little effect on any of
the metrics, while age had an effect on all four. For each topic and
age group pairing, we used each metric's version of the model (with
$g_i$ arbitrarily fixed to male) to predict the values of that metric
for query difficulties between zero and one in increments of 0.05. In
figures~\ref{fig:metpred-qu} and~\ref{fig:metpred-pcc-hard}, we depict
these values for graded utility and page click count; we show only the
six hardest query difficulties. These plots indicate that older users
have slightly higher values than younger users. In
figures~\ref{fig:metpred-scc-easy} and~\ref{fig:metpred-scc-hard} we
depict these values for successful click count; we show the six
easiest and six hardest query difficulties. Again, older users have
slightly higher values than younger users. This difference is more
pronounced for more difficult queries, suggesting that age has a
bigger effect on satisfaction for these queries.

Although we found that age had an effect on all four satisfaction
metrics, we cannot conclude that our results mean that older users are
more satisfied than younger users. Because our model only controls for
the topic of the query and the demographics of the user who issued the
query, these differences may be due to differences in intent or other
demographic-specific variation, as well as differences in user
satisfaction. Irrespective of the cause of these differences, the
contrast between these results and the results in the previous section
highlights the need for evaluation metrics that are not confounded by
demographic-specific variation.

\section{Estimating Differences}
\label{sec:pairwise-sat}

Although the methods described in sections~\ref{sec:context}
and~\ref{sec:multilevel} shed light on observed differences in
evaluation metrics between demographic groups, they do not directly
address our goal of measuring latent differences in user
satisfaction. For our third method, which we present in this section,
we therefore take a different approach. Rather than estimating
absolute levels of satisfaction for each demographic group and then
comparing these estimates, this method estimates differences in
satisfaction between demographic groups directly. First, it considers
randomly selected pairs of impressions (for the same query, issued by
users from different demographic groups) and uses a high-precision
algorithm to estimate which impression led to greater user
satisfaction. Using these labels, it then models differences in
satisfaction.


We restricted the data to only those queries that were issued by users
from at least three demographic groups and that had at least ten
impressions. We then randomly selected 10\% (roughly 62,000) of these
queries. For each query, we randomly selected 10,000 pairs of
impressions, resulting in a total of 2.7 billion pairs. Finally, for
each pair, we compared the impressions' values of the evaluation
metrics and labeled one of the impressions as leading to greater user
satisfaction if there was a difference so large that it was unlikely
to be explained by anything other than a genuine difference in user
satisfaction. By performing these preprocessing and labeling steps, we
were able to construct a high-precision--low-recall proxy for pairwise
differences in user satisfaction.

\begin{figure}
  \centering
  \begin{subfigure}{\linewidth}
  \vspace{1.5mm}
\begin{algorithmic}
\small{
\If {$\reformulationRate_i < \reformulationRate_j$} \Return $+$1
\EndIf
\If {$\reformulationRate_i > \reformulationRate_j$} \Return $-$1
\EndIf
\If {$\gradedUtility_i - \gradedUtility_j > 0.4$} \Return $+$1
\EndIf
\If {$\gradedUtility_j - \gradedUtility_i > 0.4$} \Return $-$1
\EndIf
\If {$\satifiedClickCount_i - \satifiedClickCount_j > 2$} \Return $+$1
\EndIf
\If {$\satifiedClickCount_j - \satifiedClickCount_i > 2$} \Return $-$1
\EndIf
\If {$\gradedUtility_i - \gradedUtility_j >0.2
  \textrm{ and } \satifiedClickCount_i - \satifiedClickCount_j > 1$} \Return $+$1
\EndIf
\If {$\gradedUtility_j - \gradedUtility_i > 0.2
  \textrm{ and } \satifiedClickCount_j - \satifiedClickCount_i >1$} \Return $-$1
\Else \, \Return 0
\EndIf
}
\end{algorithmic}
\caption{\label{fig:alg}Bing}
\end{subfigure}
  \begin{subfigure}{\linewidth}
      \vspace{3mm}
\begin{algorithmic}
  \small{
    \If {$\pageClickCount_i - \pageClickCount_j > 2$} \Return
    $+$1
    \EndIf
    \If {$\pageClickCount_j - \pageClickCount_i > 2$}
    \Return $-$1
    \Else\, \Return 0
    \EndIf
  }
  \caption{\label{fig:alg_2}comScore}
\end{algorithmic}
  \end{subfigure}
  \caption{Algorithms for labeling a pair of impressions.}
\vspace{-2mm}
\end{figure}

We provide the algorithm that we used to compare the impressions'
metric values in figure~\ref{fig:alg}. The metrics are ordered
according to importance. For example, reformulation rate is thought to
be a strong indicator of dissatisfaction. The algorithm therefore
considers reformulation rate first. If there is a difference, it
returns a label without considering the other metrics. If there is no
difference, it moves on to consider graded utility, followed by
successful click count. Finally, it considers graded utility and
successful click count together, using a slightly less conservative
threshold; if there are differences in both metrics, these differences
are more likely to reflect genuine differences in user satisfaction.

We obtained the thresholds using the model described in the previous
section. Specifically, we used our estimates of the effects of
demographics factors on the metrics to derive conservative upper
bounds on the effects of demographic-specific variation. For each
metric, we then used the corresponding bound to derive a minimum
threshold for differences that are so large that they are unlikely to
be explained by anything other than genuine differences in user
satisfaction. For example, if the difference in graded utility between
groups had a maximum of $\delta$, then we set the minimum threshold to
$k \delta$, where $k > 1$ reflects our confidence that two impressions
whose graded utility values differ by at least $k \delta$ correspond
to a genuine difference in user satisfaction. A higher value of $k$
yields a higher-precision--lower-recall algorithm. We set $k =2.5$ to
obtain the values in figure~\ref{fig:alg}.\looseness=-1




We used a single-level model to estimate latent differences in user
satisfaction across demographic groups. This model is similar to the
one described in the previous section, but does not include
query-specific terms because we restricted the data to pairs of
impressions for the same query. Letting $S_i - S_j$ denote the latent
difference in user satisfaction between the $i^{\textrm{th}}$ and
$j^{\textrm{th}}$ impressions, the model assumes that
\begin{align}
  \label{eqn:model-sat}
  &P(S_i - S_j > 0) = \notag{}\\ &\quad f^{-1}(\mu_0 + \gamma_{a_i} +
  \gamma_{a_j} + \gamma_{g_i} + \gamma_{g_j} + \gamma_{a_i \times g_i
    \times a_j \times g_j}),
\end{align}
where $f(\cdot)$ is a logit link function and $a_i \times g_i \times
a_j \times g_j$ denotes an interaction term. The model also assumes
that the coefficients are Gaussian distributed around
zero.

\begin{figure}[t]
  \centering
  \begin{subfigure}{.5\linewidth}
    \centering
    \includegraphics[width=1.5in]{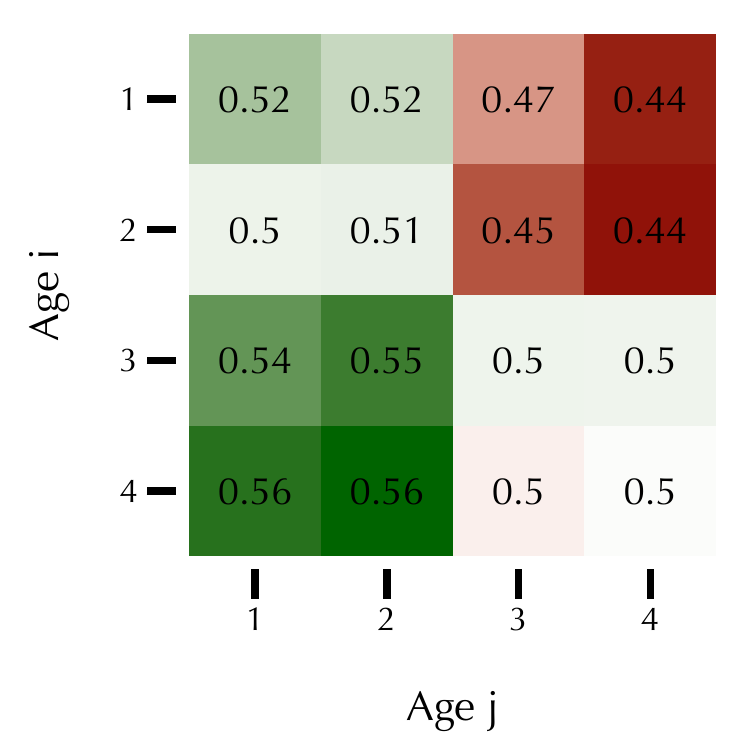}
    \caption{Bing}
    \label{fig:pairwise-bing}
    \end{subfigure}%
  \begin{subfigure}{.5\linewidth}
    \centering
  \includegraphics[width=1.5in]{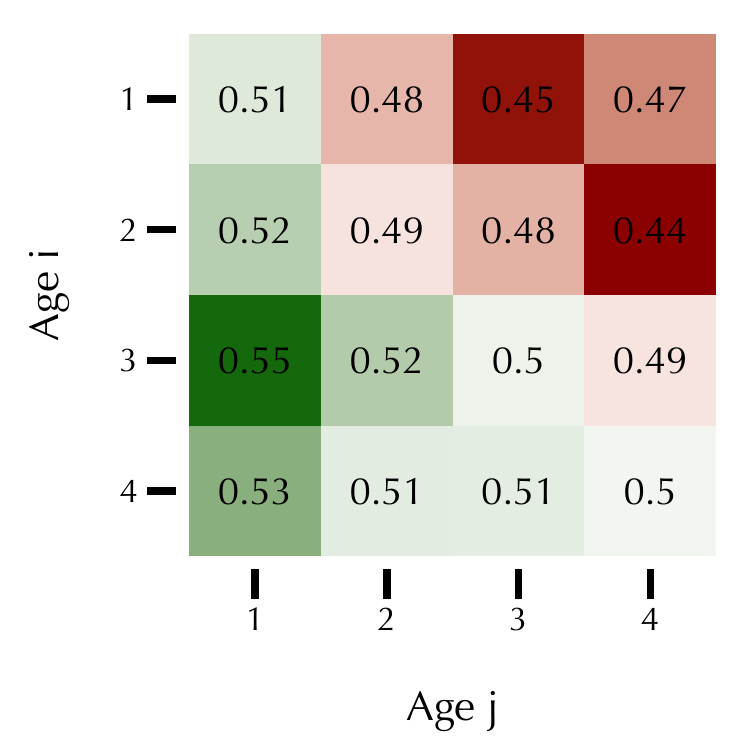}
  \caption{comScore}
  \label{fig:pairwise-comscore}
  \end{subfigure}
  \caption{$P(S_i - S_j > 0)$ for each age group pairing. Standard
    errors (not shown) are between 0.001 and 0.004.}
  \label{fig:pairwise}
\vspace{-2mm}
\end{figure}

We fit the model using pairs of impressions from different demographic
groups, labeled as either $+$1 or $-$1 via the algorithm in
figure~\ref{fig:alg}. Again, we found that gender had little
effect. To compare differences in satisfaction across age groups, we
arbitrarily fixed $g_i$ and $g_j$ to male and female,
respectively. Then, for each age group pairing, we used the model to
predict $P(S_i - S_j > 0)$. We visualize the probabilities for each
pairing in figure~\ref{fig:pairwise-bing}. This figure suggests that
older users are more satisfied than younger users, with larger
differences for users whose ages are further apart; however, because
the probabilities are all close to 0.5, the difference is relatively
small for each age group pairing. These findings are consistent with
the findings described in sections~\ref{sec:demographic}
and~\ref{sec:multilevel}; though, again, we note that these
differences may be due to other unmodeled demographic-specific
variation.

\section{External Auditing}
\label{sec:external}

To demonstrate the generality of our third method, we used this method
to conduct an external audit of a leading competitor to Bing. We used
publicly available data provided by comScore, an Internet analytics
company.\footnote{\url{http://www.comscore.com/Products/Audience-Analytics/qSearch}}
To assemble this data set, comScore recruited an opt-in consumer
panel, validated to be representative of the online population and
projectable to the total US
population~\cite{fulgoni2005professional}. The data set consists of
unfiltered search queries collected over a one week period during
November 2011. Due to a lack of detailed behavioral signals, we used
only page click count to operationalize user satisfaction. We followed
the approach described in section~\ref{sec:pairwise-sat}, again
focusing on age and gender, but rather than using the algorithm in
figure~\ref{fig:alg}, we labeled each pair of impressions using the
algorithm in figure~\ref{fig:alg_2}.

We fit the model described in the previous section (i.e.,
equation~\ref{eqn:model-sat}) using 1.2 million pairs of impressions
from different demographic groups, labeled as either $+$1 or $-$1 via
the algorithm in figure~\ref{fig:alg_2}. Again, we found that gender
had little effect, so we arbitrarily fixed $g_i$ and $g_j$ to male and
female, respectively. For each age group pairing, we then used the
model to predict $P(S_i - S_j > 0)$. We visualize the probabilities
for each pairing in figure~\ref{fig:pairwise-comscore}. Similar to the
results in the previous section, this figure suggests that older users
tend to be slightly more satisfied than younger users.

\section{Discussion}
\label{sec:discussion}

Internally auditing search engines for equal access is much more
complicated than comparing evaluation metrics for demographically
binned search impressions. In this paper, we addressed this challenge
by proposing three methods for measuring latent differences in user
satisfaction from observed differences in evaluation metrics. We then
used these methods to audit Bing, focusing specifically on age and
gender. Overall, we found no difference in satisfaction between male
and female users, but we did find that older users appear to be
slightly more satisfied than younger users. Because we used three
different methods, with complementary strengths, we can be confident
that any trends detected by all three methods are genuine, though we
cannot conclude that they were due to differences in user
satisfaction, as opposed to unmodeled demographic-specific variation.

We then used our third method to conduct an external audit of a
leading competitor to Bing, using publicly available data from
comScore. Again, we found that older users tend to be slightly more
satisfied than younger users. Because we saw the same trends for two
independently developed search engines, we hypothesize that these
trends are likely due to unmodeled differences between demographic
groups, rather than genuine differences in user satisfaction. That
said, we believe that this finding is important and should be explored
further. Graded utility and successful click count both depend on
dwell-time thresholds. Although previous laboratory experiments have
not shown substantial differences in reading times between older and
younger populations~\cite{akutsu:psychophysics-of-reading}, other work
has shown differences in reading times for scenarios involving a
mixture of relevant and non-relevant
text~\cite{connelly:distraction-age-reading-time}. Moreover, many
online services other than search engines also use evaluation metrics
based on dwell
time~\cite{hu:dwell-time-recommendation,nunez-valdez:dwell-time-ebooks,yin:dwell-time-recommendation}.

We conclude that there is a need for further investigation into
observed differences in evaluation metrics across demographic groups,
as well as a need for new metrics that are not confounded with
demographics and can be computed without using costly explicit
feedback elicitation methods.

\bibliographystyle{abbrv}
\vspace*{0.5mm}
\scriptsize
{

}
\balancecolumns
\end{document}